\documentclass[11pt]{article}
\usepackage{amsmath,amssymb}

\newcommand{\f}[2]{\frac{#1}{#2}}
\newcommand{\ko}[1]{\left( #1 \right)}

\newcommand{\kket}[1]{| #1 \rangle}

\newcommand{\bmt}[1]{{{\mbox{\boldmath$ #1 $}}}}

\newcommand{\mr}[1]{\mathrm{#1}}

\def\N{{\mathcal{N}}}

\def\cC{{\mathcal{C}}}
\def\bcC{{{\mbox{\boldmath$\mathcal C$}}}}
\def\bcP{{{\mbox{\boldmath$\mathcal P$}}}}
\def\bcK{{{\mbox{\boldmath$\mathcal K$}}}}
\def\R{{\mathbb{R}}}

\def\cP{{\mathcal{P}}}
\def\cK{{\mathcal{K}}}

\def\fC{{\mathfrak{C}}}
\def\fR{{\mathfrak{R}}}
\def\fL{{\mathfrak{L}}}
\def\fP{{\mathfrak{P}}}
\def\fK{{\mathfrak{K}}}
\def\fQ{{\mathfrak{Q}}}
\def\fS{{\mathfrak{S}}}
\def\fJ{{\mathfrak{J}}}
\def\fpsu{{\mathfrak{psu}}}
\def\fsu{{\mathfrak{su}}}

\def\eq{\equiv}
\def\be{\beta}
\def\al{\alpha}
\def\ga{\gamma}

\def\ep{{\epsilon}}
\def\half{{\mbox{$\f{1}{2}$}}}
\def\b1{{\bf{1}}}
\def\c{{\bf{c}}}

\def\A{{\bf{A}}}
\def\B{{\bf{B}}}
\def\C{{\bf{C}}}
\def\D{{\bf{D}}}
\def\a{{\bf{a}}}
\def\bb{{\bf{b}}}
\def\c{{\bf{c}}}
\def\d{{\bf{d}}}

\def\bul{{\small ~\,$\bullet$~\,}}

\def\bx{{\hspace{0.6mm}\unitlength 0.1in
\begin{picture}(1.00,1.00)(10.00,-13.00)
\special{pn 8}%
\special{pa 1000 1200}%
\special{pa 1100 1200}%
\special{pa 1100 1300}%
\special{pa 1000 1300}%
\special{pa 1000 1200}%
\special{fp}%
\end{picture}%
\hspace{0.7mm}}}
\def\sbx{{\hspace{0.6mm}\unitlength 0.1in
\begin{picture}(1.00,1.00)(10.00,-13.00)
\special{pn 8}%
\special{pa 1000 1200}%
\special{pa 1100 1200}%
\special{pa 1100 1300}%
\special{pa 1000 1300}%
\special{pa 1000 1200}%
\special{fp}%
\special{pn 8}%
\special{pa 1100 1200}%
\special{pa 1000 1300}%
\special{fp}%
\end{picture}%
\hspace{0.7mm}}}
\def\vtwosbx{{\hspace{1.0mm}\unitlength 0.1in
\begin{picture}(1.00,2.00)(16.00,-20.00)
\special{pn 8}%
\special{pa 1600 1900}%
\special{pa 1700 1900}%
\special{pa 1700 2000}%
\special{pa 1600 2000}%
\special{pa 1600 1900}%
\special{fp}%
\special{pn 8}%
\special{pa 1700 1900}%
\special{pa 1600 2000}%
\special{fp}%
\special{pn 8}%
\special{pa 1600 1800}%
\special{pa 1700 1800}%
\special{pa 1700 1900}%
\special{pa 1600 1900}%
\special{pa 1600 1800}%
\special{fp}%
\special{pn 8}%
\special{pa 1700 1800}%
\special{pa 1600 1900}%
\special{fp}%
\end{picture}
\hspace{1.0mm}}}
\def\Qbx{\,{\hspace{0mm}
\unitlength 0.1in
\begin{picture}(5.95,1.35)(19.55,-19.55)
\special{pn 8}%
\special{pa 1955 1855}%
\special{pa 2055 1855}%
\special{pa 2055 1955}%
\special{pa 1955 1955}%
\special{pa 1955 1855}%
\special{fp}%
\special{pn 8}%
\special{pa 2055 1855}%
\special{pa 2155 1855}%
\special{pa 2155 1955}%
\special{pa 2055 1955}%
\special{pa 2055 1855}%
\special{fp}%
\special{pn 8}%
\special{pa 2480 1855}%
\special{pa 2580 1855}%
\special{pa 2580 1955}%
\special{pa 2480 1955}%
\special{pa 2480 1855}%
\special{fp}%
\special{pn 8}%
\special{pa 2160 1855}%
\special{pa 2210 1855}%
\special{fp}%
\special{pn 8}%
\special{pa 2160 1955}%
\special{pa 2210 1955}%
\special{fp}%
\special{pn 8}%
\special{pa 2430 1855}%
\special{pa 2530 1855}%
\special{fp}%
\special{pn 8}%
\special{pa 2430 1955}%
\special{pa 2530 1955}%
\special{fp}%
\put(23.2800,-19.1300){\makebox(0,0){$\cdots$}}%
\end{picture}%
\hspace{0mm}}~}
\def\Qsbx{\,{\hspace{0mm}
\unitlength 0.1in
\begin{picture}(5.95,1.35)(19.55,-19.55)
\special{pn 8}%
\special{pa 2055 1855}%
\special{pa 1955 1955}%
\special{fp}%
\special{pn 8}%
\special{pa 1955 1855}%
\special{pa 2055 1855}%
\special{pa 2055 1955}%
\special{pa 1955 1955}%
\special{pa 1955 1855}%
\special{fp}%
\special{pn 8}%
\special{pa 2055 1855}%
\special{pa 2155 1855}%
\special{pa 2155 1955}%
\special{pa 2055 1955}%
\special{pa 2055 1855}%
\special{fp}%
\special{pn 8}%
\special{pa 2155 1855}%
\special{pa 2055 1955}%
\special{fp}%
\special{pn 8}%
\special{pa 2480 1855}%
\special{pa 2580 1855}%
\special{pa 2580 1955}%
\special{pa 2480 1955}%
\special{pa 2480 1855}%
\special{fp}%
\special{pn 8}%
\special{pa 2155 1955}%
\special{pa 2205 1905}%
\special{fp}%
\special{pn 8}%
\special{pa 2580 1855}%
\special{pa 2480 1955}%
\special{fp}%
\special{pn 8}%
\special{pa 2160 1855}%
\special{pa 2210 1855}%
\special{fp}%
\special{pn 8}%
\special{pa 2160 1955}%
\special{pa 2210 1955}%
\special{fp}%
\special{pn 8}%
\special{pa 2430 1855}%
\special{pa 2530 1855}%
\special{fp}%
\special{pn 8}%
\special{pa 2430 1955}%
\special{pa 2530 1955}%
\special{fp}%
\put(23.2800,-19.1300){\makebox(0,0){$\cdots$}}%
\special{pn 8}%
\special{pa 2430 1905}%
\special{pa 2480 1855}%
\special{fp}%
\end{picture}%
\hspace{0mm}}~}
\def\twobx{{\bx\hspace{-1.3mm}\bx}}
\def\twosbx{{\sbx\hspace{-1.3mm}\sbx}}
\begin{document}
\quad 
\vspace{-1.5cm}

\begin{flushright}
\parbox{3cm}
{
{\bf October 2006} \\ 
DAMTP-06-$64$ \hfill \\
UT-06-17 \hfill \\
{\tt hep-th/0610295}\hfill
 }
\end{flushright}

\vspace*{0.5cm}

\begin{center}
\Large\bf 
The Asymptotic Spectrum of the \bmt{\mathcal{N}=4} \\ 
Super Yang-Mills Spin Chain
\end{center}
\vspace*{0.7cm}
\centerline{\large 
Heng-Yu Chen$^{\dagger,\, a}$\,,~
Nick Dorey$^{\dagger,\, b}$
~and~
Keisuke Okamura$^{\ddagger,\, c}$}
\vspace*{0.5cm}
\begin{center}
${}^{\dagger}$\emph{DAMTP, Centre for Mathematical Sciences, Cambridge University,\\
Wilberforce Road, Cambridge CB3 OWA, UK.} \\
\vspace{0.3cm}
~and~\\
\vspace{0.3cm}
${}^{\ddagger}$\emph{Department of Physics, Faculty of Science, 
University of Tokyo,\\
Bunkyo-ku, Tokyo 113-0033, Japan.} \\
\vspace*{0.5cm}
%E-mail:
${}^{a,\, b}$ {\tt h.y.chen, n.dorey@damtp.cam.ac.uk}\\%\quad 
${}^{c}$ {\tt okamura@hep-th.phys.s.u-tokyo.ac.jp}
\end{center}

\vspace*{0.7cm}

\centerline{\bf Abstract} 

\vspace*{0.5cm}

In this paper we discuss the asymptotic spectrum of the spin chain
description of planar ${\cal N}=4$ SUSY Yang-Mills. The states
appearing in the spectrum belong to irreducible representations 
of the unbroken supersymmetry 
$SU(2|2)\times SU(2|2)$ with non-trivial extra central extensions. 
The elementary
magnon corresponds to the bifundamental representation while
boundstates of $Q$ magnons form a certain short representation
of dimension $16Q^{2}$. Generalising the Beisert's analysis of the
$Q=1$ case, we derive the exact dispersion relation for these states
by purely group theoretic means.

\vspace*{1.0cm}

\vfill

\thispagestyle{empty}
\setcounter{page}{0}
\setcounter{footnote}{0}

\newpage 
%%%%%%%%%%%%%%%%%%%%%%%%%%%%%%%%%%%%%%%%%
%%%%%%%%%%%%%%%%%%%%%%%%%%%%%%%%%%%%%%%%%
%%%%%%%%%%%%%%%%%%%%

The emergence of integrability on both sides of the AdS/CFT
correspondence \cite{MZ,B,BPR} continues to provide improvements in our
understanding of large-$N$ gauge theory and string theory. 
Recent progress has centered on a particular limit
\cite{Hofman:2006xt} 
where the spin chain describing the single trace operators of ${\cal N}=4$ SUSY
Yang-Mills theory becomes infinitely long.\footnote{The importance of this limit was also stressed earlier 
in \cite{Staudacher:2004tk}.}
Specifically one considers
a limit where the $U(1)$ R-charge $J_{1}$ and 
scaling dimension $\Delta$ of the operator become large 
with the difference $E=\Delta-J_{1}$ and the 't Hooft coupling
$\lambda$ held fixed. In this limit, the spectrum corresponds to
localised excitations which propagate almost freely on the infinite
chain. The remaining interactions between these excitations are governed by
a factorisable S-matrix. In this paper we will describe 
the minimal possibility for the complete 
spectrum of asymptotic states of the spin
chain.\footnote{The issue of completeness will be discussed further below.}

The asymptotic states mentioned above correspond to local excitations
above the ferromagnetic groundstate of the spin chain. The latter 
state corresponds
to the gauge theory operator ${\mr{Tr}}\left(Z^{J_{1}}\right)$ where
$Z$ is a complex adjoint scalar field with R-charge $J_{1}=1$. The
ferromagnetic groundstate is
not invariant under the full superconformal algebra $\fpsu(2,2|4)$,
but instead is only preserved by the subalgebra 
$\left(\fpsu(2|2)\times \fpsu(2|2)\right)\ltimes {\mathbb{R}}$. 
The residual symmetry algebra can also be understood as two copies
of $\fsu(2|2)$ with their central charges identified.
This common central charge will play the role of Hamiltonian for 
the associated spin chain 
whose eigenvalue is identified with the combination $\Delta-J_{1}$.
Moreover as noted in \cite{Beisert:2005tm}, 
an important subtlety arising is that this symmetry
algebra needs to be further extended 
by two additional central charges in order to 
describe excitations of non-zero momenta.
This extended unbroken symmetry 
is linearly realised on excitations above the groundstate 
which consequently form representations of the corresponding
non-abelian symmetry group 
$(PSU(2|2)\times PSU(2|2))\ltimes {\mathbb{R}}^{3}$. 
In the following we will determine 
which representations appear in the spectrum of asymptotic states. 

The fundamental excitation of the spin chain, known as the magnon, 
corresponds to an insertion of a single impurity, with definite
momentum $p$, into the 
groundstate operator ${\mr{Tr}}\left(Z^{J_{1}}\right)$.  
There are a total of sixteen possible choices for the impurity
corresponding to the various scalars and spinor fields and covariant
derivatives of the ${\cal N}=4$ theory \cite{Berenstein:2002jq}. 
As we review below, these excitations fill out a multiplet 
in the bifundamental representation of
$(PSU(2|2)\times PSU(2|2))\ltimes {\mathbb{R}}^{3}$. In terms of the centrally-extended algebra
described above, these are short representations with an exact BPS
dispersion relation which is uniquely given by the closure of the
algebra to be \cite{Beisert:2005tm,BDS,hub},   
\begin{equation}
E=\Delta-J_{1}=\sqrt{1+8g^{2}\sin^{2}\left(\frac{p}{2}\right)}\,.
\label{Magnondispersion}
\end{equation}
Here, following the convention of \cite{Beisert:2005tm}, we have introduced a coupling 
$g$ which is related to the 
't Hooft coupling $\lambda$ by $g^{2}=\lambda/8\pi^{2}$. As the
residual symmetry generators commute with the Hamiltonian of the spin chain,
each state in the multiplet has the same dispersion relation 
(\ref{Magnondispersion}). With
this in mind, we can think of the sixteen states in the bifundamental 
multiplet as distinct ``polarisations'' of a single excitation.   

The full spin chain for the planar ${\cal N}=4$ theory has a closed
subsector, known as the $SU(2)$ sector, 
where only impurities corresponding to one complex adjoint
scalar field are included. Equivalently, we restrict our attention to magnons
of a single polarisation. Within this subsector, it is known that the
asymptotic spectrum also includes an infinite tower of magnon
boundstates \cite{Dorey:2006dq}. These excitations are labelled by a
positive integer $Q$, which corresponds to the number of constituent
magnons of different flavours, as well as their conserved momentum $p$. The location of the
corresponding poles in the exact magnon S-matrix indicates that these
states have an exact dispersion relation of the form,  
\begin{equation}
E=\Delta-J_{1}=\sqrt{Q^{2}+8g^{2}\sin^{2}\left(\frac{p}{2}\right)}
\label{Magnondispersion2}
\end{equation}
which generalises (\ref{Magnondispersion}). 
The corresponding classical string solution which 
precisely reproduces (\ref{Magnondispersion2})
has been found in \cite{Chen:2006ge,Minahan:2006bd}. Scattering matrices for these states 
have recently also been constructed in \cite{Chen:2006gq,Roiban:2006gs}.
In the context of the full
model, these asymptotic states in the $SU(2)$ sector should be particular
representatives from complete representations of the symmetry group 
$(PSU(2|2)\times PSU(2|2))\ltimes {\mathbb{R}}^{3}$. In fact, we will 
see below that the
$Q$-magnon boundstate lies in a short irreducible representation of
dimension $16Q^{2}$ \cite{Dtalk,Beisert3}.
 The representation in question can be thought of
as a supersymmetric extension of the rank-$Q$ traceless symmetric tensor
representation of the unbroken $SO(4)\simeq SU(2)\times SU(2)$
R-symmetry which is a subgroup of $(PSU(2|2)\times PSU(2|2))
\ltimes{\mathbb{R}}^{3}$. This particular representation includes the 
known BPS boundstates of 
magnons in the $SU(2)$ sector. 
An important consistancy check is that the representation 
does not lead to boundstates in any of the other rank one subsectors which 
are known to be absent \cite{Hofman:2006xt}. 
In \cite{Beisert:2005tm}, the dispersion relation
(\ref{Magnondispersion}) for excitations transforming in the bifundamental
representation of $(PSU(2|2)\times PSU(2|2))\ltimes {\mathbb{R}}^{3}$ 
was derived from purely group theoretical means. 
As an additional test of our results, 
we will extend the analysis to the symmetric
tensor representations relevant for the boundstates described above 
to provide a parallel group theoretic derivation of the dispersion
relation (\ref{Magnondispersion2}).     
         
To begin, let us first focus on a single copy of $\fsu (2|2)\subset
\left(\fpsu(2|2)\times \fpsu(2|2)\right)\ltimes {\mathbb{R}}$ 
and review some associated basic facts
following \cite{Beisert:2005tm}. The algebra consists of two bosonic generators
$\fL^{\al}{}_{\be}$ and $\fR^{a}{}_{b}$ which generate
$\fsu(2)\times\fsu(2)$ 
rotations;
two fermionic supersymmetry generators $\fQ^{\al}{}_{b}$ and
$\fS^{a}{}_{\be}$, and finally the algebra 
also contains a central charge $\fC$ which is shared with the other $\fsu(2|2)$.
These generators obey the following (anti-)commutation relations:
\begin{align}
[\fR^{a}{}_{b},
  \fJ^{c}]&=\delta^{c}_{b}\fJ^{a}-\half\delta^{a}_{b}\fJ^{c}\,,
\label{comm1}\\
[\fL^{\al}{}_{\be},
  \fJ^{\ga}]&=\delta^{\ga}_{\be}\fJ^{\al}-\half\delta^{\al}_{\be}\fJ^{\ga}\,,
\label{comm2}\\
\{ \fQ^{\al}{}_{a}, \fS^{b}{}_{\be}
\}&=\delta^{b}_{a}\fL^{\al}{}_{\be}+\delta^{\al}_{\be}\fR^{b}{}_{a}+
\delta^{b}_{a}
\delta^{\al}_{\be}\fC\,,
\label{comm3}
\end{align}
where $\fJ$ stands for any generator with appropriate indices.

In addition, as discussed in \cite{Beisert:2005tm}, the $\fsu(2|2)$
algebra is too restrictive for the discussion of excitations with
non-zero momentum and it is
necessary to enlarge it to $\fsu(2|2)\ltimes
\R^{2}\cong\fpsu(2|2)\ltimes {\mathbb{R}}^{3}$, with two extra 
central charges $\fP$ and $\fK$ satisfying the 
anti-commutation relations, 
\begin{alignat}{3}
\{ \fQ^{\al}{}_{a}, \fQ^{\be}{}_{b} \}&=\ep^{\al\be}\ep_{ab}\fP\,,\qquad 
&\{ \dot\fQ^{\dot\al}{}_{\dot a}, \dot\fQ^{\dot\be}{}_{\dot b}
\}&=\ep^{\dot \al\dot \be}\ep_{\dot a\dot b}\fP\,,
\label{comm4}\\
\{ \fS^{a}{}_{\al}, \fS^{b}{}_{\be} \}&=\ep^{ab}\ep_{\al\be}\fK\,,\qquad 
&\{ \dot\fS^{\dot a}{}_{\dot \al}, \dot\fS^{\dot b}{}_{\dot \be} \}&=
\ep^{\dot a\dot b}\ep_{\dot \al\dot \be}\fK\label{comm5}\,.
\end{alignat}
The two extra central charges $\fP$ and $\fK$ are unphysical in the
sense that they vanish when the constraint of vanishing total momentum 
is imposed.\footnote{These two extra central charges $\fP$ and $\fK$ 
in fact combine with $\fC$ to give a vector under group $SO(1,2)$ \cite{Beisert:2005tm,Hofman:2006xt}.}
The full extended subalgebra is then obtained by taking direct product
between two copies of $\fpsu(2|2)\ltimes{\mathbb{R}}^{3}$ and
identifying their central charges (both physical and unphysical ones),
which extend the residual symmetry algebra from
$\left(\fpsu(2|2)\times\fpsu(2|2)\right)\ltimes{\mathbb{R}}$ 
to $\left(\fpsu(2|2)\times\fpsu(2|2)\right)\ltimes{\mathbb{R}}^{3}$.
Under the extended residual symmetry algebra, the central charge $\fC$
can be identified with the Hamiltonian for the spin chain, whereas the two extra
central charges play the role of gauge transformation generators 
which insert or remove a background chiral field $Z$ \cite{Beisert:2005tm}.
 
The fundamental representation of $\fpsu(2|2)\ltimes {\mathbb{R}}^{3}\simeq\fsu(2|2)\ltimes{\mathbb{R}}^{2}$ corresponds to a $2|2$ dimensional superspace given by the basis
\begin{equation}
\sbx \equiv\left(\begin{array}{c}\phi^{a}\\
  \psi^{\alpha}\end{array}\right)
\,,~~~ a=1\,,2\,,\quad \alpha=1\,,2\,.
\end{equation}
Here we have adopted the notation for super Young tableau introduced in 
\cite{BahaBalantekin:1980pp}. The fields $\phi^{a}$ and
$\psi^{\alpha}$ are bosonic and fermionic, respectively.  
The group generators acting on this space can be written in the following 
$4\times 4$ supermatrix form:
\begin{equation} 
\left(
\begin{array}{c|c}
\fR^{a}{}_{b} & \fQ^{\al}{}_{b}\\
\hline
\fS^{a}{}_{\be} & \fL^{\alpha}{}_{\beta}\\
\end{array}
 \right)
\end{equation}
where $\fR$ and $\fL$ are $2\times 2$ hermitian $SU(2)$
 generators, whereas the entries in $\fQ$ and $\fS$ are complex
 Grassmann variables. We can decompose the fundamental representation $\sbx$ under the maximal 
bosonic subgroup $SU(2)\times SU(2)$ as, 
\begin{equation}
\sbx =\overbrace{(\bx, {\bf{1}})}^{\mbox{$\vphantom{\f{}{}}\phi^{a}$}}\oplus
\overbrace{({\bf{1}}, \bx)}^{\mbox{$\vphantom{\f{}{}}\psi^{\alpha}$}}\,.\label{elemag1}
\end{equation} 

The canonical action of the $\fpsu(2|2)\ltimes {\mathbb{R}}^{3}$ generators on the components $\phi^{a}$ and $\psi^{\alpha}$ are then given by \cite{Beisert:2005tm}
\begin{align}
\fQ^{\al}{}_{a}\kket{\phi^{b}}&=\a\,\delta^{b}_{a}\kket{\psi^{\al}}\,,
\label{Q-phi}\\
\fQ^{\al}{}_{a}\kket{\psi^{\be}}&=\bb\,\ep^{\al\be}\ep_{ab}\kket{\phi^{b}Z^{+}}\,,
\label{Q-psi}\\
\fS^{a}{}_{\al}\kket{\phi^{b}}&=\c\,\ep^{ab}\ep_{\al\be}\kket{\psi^{\be}Z^{-}}\,,
\label{S-phi}\\
\fS^{a}{}_{\al}\kket{\psi^{\be}}&=\d\,\delta^{\be}_{\al}\kket{\phi^{a}}
\label{S-psi}\,,
\end{align}
whereas the $SU(2)$ generators $\fR$ and $\fL$ act on bosonic and fermionic components as
\begin{align}
\fR^{a}{}_{b}\kket{\phi^{c}}&=\delta^{c}_{b}\kket{\phi^{a}}-\half \delta^{a}_{b}\kket{\phi^{c}}\,,
\label{R-phi}\\
\fL^{\al}{}_{\be}\kket{\psi^{\ga}}&=\delta^{\ga}_{\be}\kket{\psi^{\al}}-\half \delta^{\al}_{\be}\kket{\psi^{\ga}}\,.
\label{L-phi}
\end{align}
Here $\a$, $\bb$, $\c$ and $\d$ can be expressed as functions of the magnon spectral parameters $x^{+}$ and $x^{-}$, which in turns are related to individual magnon momentum $p$ by
\begin{equation}
\exp(ip)=\frac{x^{+}}{x^{-}}\,.\label{momentum}
\end{equation}
The symbols $Z^{\pm}$ in (\ref{Q-psi}) and (\ref{S-phi}) denote an inserting $(+)$ or a removing $(-)$ of a background $Z$ field on the right of the excitation $\phi^{a}$ or $\psi^{\alpha}$, respectively.
It is important to note that the fundamental representation $\sbx$ 
is in fact a short or atypical 
representation of $\fpsu(2|2)\ltimes {\mathbb{R}}^{3}$, and
it satisfies the shortening condition which for this case 
is given in terms of the three central charges $\cC$, $\cP$ and $\cK$ (eigenvalues of $\fC$, $\fP$ and $\fK$, respectively) as \cite{Beisert3,BeisertTalk}
\begin{equation}
\cC^{2}-\cP\cK=\frac{1}{4}\,.\label{shortening1}
\end{equation}
Using the explicit parameterisations for the central charges in terms of spectral parameters given in \cite{Beisert:2005tm}, the shortening condition is equivalent to the constraint on the magnon spectral parameters:
\begin{equation}
x^{+}+\frac{g^{2}}{2x^{+}}-x^{-}-\frac{g^{2}}{2x^{-}}=i\,.\label{constraint1}
\end{equation}
The exact magnon dispersion relation (\ref{Magnondispersion}) then arises from the protected central charge 
$\cC$ carried by the fundamental representation $\sbx$.

Let us recall here that, in terms of ${\cal N}=4$ SUSY Yang-Mills, 
the elementary excitation of the spin chain corresponds to the
insertion of an impurity field with\footnote{Here $\Delta_{0}$ denotes
  the bare dimension of the inserted field.} $\Delta_{0}-J_{1}=1$ 
into $\mathrm{Tr}\left(Z^{J_{1}}\right)$. In the limit
$J_{1}\rightarrow \infty$, this corresponds to a single 
magnon propagating over the ferromagnetic groundstate of the infinite chain.
There are eight bosonic and eight fermionic impurities which
correspond to sixteen different possible polarisations of the magnon. 
Explicitly, they correspond to different elements of the set  
$\{ \Phi_{i}, D_{\mu}, \Psi_{\al \be}, \Psi_{\dot\al\dot \be} \}$. Here 
$i$, $\mu=1,\dots,4$ are indices in the vector representation of the 
two $SO(4)$ factors left unbroken by the ferromagnetic groundstate. 
The former is the unbroken R-symmetry of the ${\cal N}=4$ theory while 
the latter corresponds to conformal spin. In view of their
interpretation as rotations in the dual string geometry, we denote these 
$SO(4)_{\rm S^{5}}$ and $SO(4)_{\rm AdS_{5}}$, respectively. 
The scalars $\Phi_{i}$ and covariant derivatives $D_{\mu}$ form a 
vector representation of each group. We also use the standard isomorphism 
$SO(4)\simeq SU(2)_{\rm L}\times SU(2)_{\rm R}$ to introduce dotted
and undotted spinor indices for each factor. The fermionic fields of
the ${\cal N}=4$ theory, denoted $\Psi_{\al \be}$, $\Psi_{\dot\al\dot
  \be}$ $(\al, \dot\al=1,2)$ transform in the appropriate bispinor representations.  
The quantum numbers of the ${\cal N}=4$ fields under the bosonic
symmetries are summarised in the following table (for more details, see for example \cite{Sadri:2003pr}).
\begin{table}[htdp]
\caption{$SU(2)^{4}$ representations of $\N=4$ fields.}
\begin{center}
\begin{tabular}{l|lcccccccr|c|c}
Fields	&	& $\!\!\!\!SU(2)_{{\rm S}^{5},{\rm L}}\!\!\!\!$&$\!\!\!\!
\times\!\!\!\!$&$\!\!\!\!SU(2)_{{\rm AdS}_{5},{\rm R}}
\!\!\!\!$&$\!\!\!\!\times\!\!\!\!$&$\!\!\!\!SU(2)_{{\rm S}^{5},{\rm R}}
\!\!\!\!$&$\!\!\!\!\times\!\!\!\!$&$\!\!\!\!
SU(2)_{{\rm AdS}_{5},{\rm L}}\!\!\!\!$
& &	$\Delta_{0}-J_{1}$	&	$\Delta_{0}+J_{1}$\\
\hline\hline
\, $Z$	&	$($&$ {\bf 1}$&$,$&${\bf 1}$&$;$&${\bf 1}$&$,$&${\bf 1}$&$)$	&	$0$	&$2$\\
\, $\bar Z$	&	$($&$ {\bf 1}$&$,$&${\bf 1}$&$;$&${\bf 1}$&$,$&${\bf 1}$&$)$	&	$2$	&$0$\\
\, $\Phi_{i}$	&	$($&$ \bx$&$,$&$\b1$&$;$&$\bx$&$,$&$\b1$&$)$	&	$1$	&$1$\\
\, $D_{\mu}$	&	$($&$ {\bf 1}$&$,$&$\bx$&$;$&$\b1$&$,$&$\bx$&$)$	&	$1$	&$1$\\
\, $\Psi_{\al\be}$	&	$($&$ \bx$&$,$&$\b1$&$;$&$\b1$&$,$&$\bx$&$)$	&	$1$	&$2$\\
\, $\Psi_{{\dot\al}{\dot\be}}$	&	$($&$ {\bf 1}$&$,$&$\bx$&$;$&$\bx$&$,$&$\b1$&$)$	&	$1$	&$2$\\
%%\, $\Psi_{\al{\dot\be}}$	&	$($&$ \bx$&$,$&${\bf 1}$&$;$&${\bf %%1}$&$,$&$\bx$&$)$	&	$2$	&$1$\\
%%\, $\Psi_{{\dot\al}\be}$	&	$($&$ {\bf 1}$&$,$&$\bx$&$;$&$\bx$&$,$&${\bf %%1}$&$)$	&	$2$	&$1$\\
\end{tabular}
\end{center}
\label{tab1}
\end{table}
\newline
To interpret the impurities described above in terms of the supergroup
$(PSU(2|2)\times PSU(2|2))\ltimes{\mathbb{R}}^{3}$, we note that the bifundamental representation is 
given by the direct product between two copies of fundamental $\sbx$
described above,   
\begin{equation}
\left(\sbx;\sbx\right)=\left(\bx,\b1;\bx,\b1\right)\oplus
\left(\bx,\b1;\b1,\bx\right)\oplus
\left(\b1,\bx;\bx,\b1\right)
\oplus\left(\b1,\bx;\b1,\bx\right)\,.\label{elemag2}
\end{equation}  
Here we have also decomposed $\left(\sbx;\sbx\right)$ in terms of  
representations of the $SU(2)^{4}$ bosonic subgroup of
$(PSU(2|2)\times PSU(2|2))\ltimes {\mathbb{R}}^{3}$. 
There are again sixteen components within this decomposition,
precisely what one needs to incoporate the elementary excitations
listed in Table \ref{tab1}. By identifying the four $SU(2)$
factors in (\ref{elemag2}), column by column, with the other four in
Table \ref{tab1}, we can identify each term in (\ref{elemag2}) with an
impurity $\N=4$ theory according to,  
%\begin{alignat}{3}
%\Phi_{i}&\equiv \left(\bx,\b1;\bx,\b1\right)\,,&\qquad 
%D_{a}&\equiv \left(\b1,\bx;\b1,\bx\right)\,,\cr
%\Psi_{\al\be}&\equiv\left(\bx,\b1;\b1,\bx\right)\,,&\qquad 
%\Psi_{\dot\al\dot\be}&\equiv\left(\b1,\bx;\bx,\b1\right)\,.\label{id4}
%\end{alignat}
\begin{equation}
\Phi_{i}\equiv \left(\bx,\b1;\bx,\b1\right)\,,\quad 
D_{\mu}\equiv \left(\b1,\bx;\b1,\bx\right)\,,\quad 
\Psi_{\al\be}\equiv\left(\bx,\b1;\b1,\bx\right)\,,\quad 
\Psi_{\dot\al\dot\be}\equiv\left(\b1,\bx;\bx,\b1\right)\,.\label{id4}
\end{equation}
So the sixteen elementary excitations completely fill up the bifundamental 
representation of $SU(2|2)\times SU(2|2)$. 

Having treated the case of the elementary magnon, 
we now proceed to determine the corresponding representations of
$(\fpsu(2|2)\times \fpsu(2|2))\ltimes {\mathbb{R}}^{3}$ relevant for the magnon boundstates discovered
in \cite{Dorey:2006dq}. The natural starting point for the $Q$-magnon boundstate 
is to consider the tensor product between $Q$ copies of the 
elementary magnon representation $(\sbx;\sbx)$ as given in
(\ref{elemag2}).
In particular the magnon boundstates should transform in 
the short irreducible representations under the residual symmetry algebra $(\fpsu(2|2)\times\fpsu(2|2))\ltimes {\mathbb{R}}^{3}$.

As above we will begin by considering a single copy of $\fpsu(2|2)\ltimes{\mathbb{R}}^{3}$ 
and will start with the simplest case taking the tensor product between two 
fundamentals $\sbx$ as described in (\ref{elemag1}).
In the usual experience of dealing with Lie algebra, 
one expects that tensoring two or more irreducible representations (e.g., the fundamental representation) would
yield direct sum of irreducible representations (including long and short).
However, as pointed out in \cite{Beisert3,BeisertTalk}, such multiplet splitting does not happen generally for  
$\fpsu(2|2)\ltimes{\mathbb{R}}^{3}$. In particular, for the tensor product of two fundamental representations, 
the splitting into irreducible representations of lower dimensions 
can only happen if the central charges carried by the two constituent magnons satisfy the ``splitting condition''  
\begin{equation}
(\cC_{1}+\cC_{2})^{2}-(\cP_{1}+\cP_{2})(\cK_{1}+\cK_{2})=1\quad 
\Rightarrow \quad 
2\cC_{1}\cC_{2}-\cP_{1}\cK_{2}-\cK_{1}\cP_{2}=\frac{1}{2}\,.\label{splitting1}
\end{equation}
Here $\cC_{i}$, $\cP_{i}$ and $\cK_{i}$ are the central charges carried by the constituent magnons $i=1,2$. 
Clearly for arbitrary combinations of the central charges, (\ref{splitting1}) would not be satisfied, 
hence tensoring two fundamental representations generically gives us a long irreducible representation of sixteen dimensions. 

Interestingly, the splitting condition (\ref{splitting1}) can be satisfied when the spectral parameters obeys the boundstate pole condition established in \cite{Dorey:2006dq,Chen:2006gq}, that is 
\begin{equation}
x^{-}_{1}=x^{+}_{2}\,.\label{polecondition}
\end{equation}
This can be shown by explicitly calculating the expression in (\ref{splitting1}) using the spectral parameters.

In this special case, the long multiplet of sixteen dimensions splits into direct sum 
of two short representations of eight dimensions, and
we can label them using the branching rules for 
super Young tableaux worked out in \cite{BahaBalantekin:1980pp}, 
\begin{equation}
\sbx\otimes\sbx = \twosbx \oplus\vtwosbx\,.\label{ten2sbx}
\end{equation}
The two terms on the RHS of (\ref{ten2sbx}) represent  
distinct irreducible representations of $\fpsu(2|2)\ltimes{\mathbb{R}}^{3}$.
The first irreducible representation, denoted $\twosbx$, corresponds to a 
symmetrisation of indices for
the bosonic components $\phi^{a}$s of each fundamental representation 
and anti-symmetrisation of indices
for the corresponding Grassmann components $\psi^{\al}$s. We will call
this the ``super-symmetric'' representation. In contrast, the second
term $\vtwosbx$ corresponds to a 
``super-anti-symmetric'' representation where the bosonic/fermionic
indices are antisymmetrised/symmetrised, respectively.
Both of them are in fact short irreducible representations of
$\fpsu(2|2)\ltimes {\mathbb{R}}^{3}$, 
satisfying the shortening condition given in \cite{Beisert3} and carrying the protected central charges.

We can further decompose these short representations into representations 
under its $SU(2)\times SU(2)$ bosonic subgroup\footnote{In fact, our situation is further simplified as
the subgroups only involve $SU(2)$s, 
whose Young tableaux only contain single rows at most.}. In terms of standard 
$SU(2)$ Young tableaux the decompositions are  
\begin{align}
\twosbx & =(\twobx, {\bf 1})\oplus(\bx, \bx)\oplus({\bf 1}, {\bf 1})\,,
\label{twosbx}\\
\vtwosbx & =({\bf 1}, {\bf 1})\oplus(\bx, \bx)\oplus({\bf 1}, \twobx)\,.
\label{vtwosbx}
\end{align}

The generalisation to the physical case with two factors of $\fpsu(2|2)\ltimes {\mathbb{R}}^{3}$ with their central charges identified is straightforward. Combining (\ref{elemag2}) and (\ref{ten2sbx}), the 
tensor product of two bifundamental representations can be decomposed
as
\begin{equation}
(\sbx;\sbx) \otimes (\sbx;\sbx) = (\twosbx;\twosbx) \oplus
  (\twosbx;\vtwosbx) \oplus (\vtwosbx;\twosbx) 
\oplus (\vtwosbx;\vtwosbx)\,.
\label{sbx2-x-sbx2}
\end{equation}
Each irreducible representation in the decomposition in (\ref{sbx2-x-sbx2}) is
manifestly supersymmetric, containing equal number of bosonic and 
fermionic components. To identify the nature of the corresponding
states, it is convenient to further decompose each term in the
decomposition (\ref{sbx2-x-sbx2}) into the irreducible representations
of the four $SU(2)$ subgroups. For example, the first term 
yields, 
\begin{align}
(\twosbx;\twosbx) 
&= (\twobx,\b1;\twobx,\b1)\oplus (\twobx,\b1;\b1,\b1)\oplus(\b1,\b1;\twobx,\b1)\oplus(\b1,\b1;\b1,\b1)\cr
&\qquad {}\oplus(\bx,\bx;\b1,\b1)\oplus (\bx,\bx;\twobx,\b1)\cr
&\qquad {}\oplus(\b1,\b1;\bx,\bx)\oplus (\twobx,\b1;\bx,\bx)\cr
&\qquad {}\oplus(\bx,\bx;\bx,\bx)\,.\label{decten2bx}
\end{align}
As each state in the constituent bifundamental multiplet corresponds to an
insertion of a particular impurity in the ${\cal N}=4$ theory, we can identify the terms on the RHS of
(\ref{decten2bx}) with appropriate bilinears in the ${\cal N}=4$ fields.   
In the Appendix, we have listed the $SU(2)^{4}$ quantum numbers of
arising from each product of two ${\cal N}=4$ impurities.   
Comparing (\ref{decten2bx}) with the results in the Appendix, we
identify the relevant bilinears as, 
\begin{equation}
(\twosbx;\twosbx)\equiv ~(\Phi_{i}\otimes
  \Phi_{j})~\oplus~(\Phi_{i}\otimes \Psi_{\al\be})~ \oplus
  ~(\Phi_{i}\otimes \Psi_{\dot\al\dot\be})~ 
\oplus~(D_{a}\otimes\Phi_{i})~\,.\label{2N4fields}
\end{equation}
where appropriate (anti-)symmetrisations over indices is understood. 

As explained above, the two magnon boundstates in the $SU(2)$ sector 
must correspond to (at least) one of the short representations
of $(\fpsu(2|2)\times \fpsu(2|2))\ltimes{\mathbb{R}}^{3}$ appearing in the decomposition
(\ref{decten2bx}). To identify the relevant representation we note
that each magnon of the $SU(2)$ sector carries one unit of a second
$U(1)$ R-charge denoted $J_{2}$ in \cite{Dorey:2006dq}. The charge $J_{2}$
corresponds to one Cartan generator of the unbroken R-symmetry group 
$SO(4)\simeq SU(2)\times SU(2)\subset (PSU(2|2)\times PSU(2|2))\ltimes {\mathbb{R}}^{3}$
normalised to that states in the bifundamental representation of
$SU(2)\times SU(2)$ have charges $-1\leq J_{2}\leq 1$. The two-magnon
boundstate has charge $J_{2}=2$. It is straightforward to check that
this value is realised in the term 
$(\twobx,1;\twobx,1)$ appearing in the decomposition (\ref{decten2bx})
of the ``bi-super-symmetrised'' representation $(\twosbx;\twosbx)$ of 
$(\fpsu(2|2)\times \fpsu(2|2))\ltimes {\mathbb{R}}^{3}$. 
One may also check that the remaining irreducible representations in the decomposition 
(\ref{sbx2-x-sbx2}) of the tensor product do not contain states with 
$J_{2}=2$. 

Summarising the above discussion we deduce that the two magnon
boundstate discovered in \cite{Dorey:2006dq} is one component of a
multiplet of states in the $(\twosbx;\twosbx)$ of 
$(\fpsu(2|2)\times \fpsu(2|2))\ltimes {\mathbb{R}}^{3}$. The dimension of this representation is
sixty-four, which corresponds to the number of independent
polarisations of the two magnon boundstate. The various bilinear impurities
corresponding to these polarisations appear in (\ref{2N4fields}).   
A check on the identification described above is that there are no
bilinears involving only either two fermions or two derivatives. 
This agrees with the known absence of two magnon boundstates in the 
$SU(1|1)$ and $SL(2,{\mathbb{R}})$ sectors, respectively 
\cite{Hofman:2006xt,Minahan:2006bd,Roiban:2006gs}.

It is straightforward to extend the discussion to the case of general 
$Q$-magnon scattering, now the multiplet splitting condition can be given by
\begin{equation}
\bcC_{Q}^{2}-\bcP_{Q}\bcK_{Q}=\frac{Q^{2}}{4}\,,\label{splitting2}
\end{equation}
where $\bcC_{Q}$, $\bcP_{Q}$ and $\bcK_{Q}$ are the central charges
carried by the generic long irreducible representation formed by tensor product between $Q$ fundamentals. 
This can be satisfied when we impose the boundstate condition
\begin{equation}
x_{i}^{-}=x_{i+1}^{+}\,,~~~i=1\,,2\,,\dots\,,Q-1\,.\label{polecond2}
\end{equation}
The tensor product between $Q$ fundamental representations generally consists of 
direct sum of long representations \cite{Beisert3}. In this special limit (\ref{polecond2}), it
can be further decomposed into direct sum of short representations 
and labelled by the branching rules in \cite{BPR} as 
\begin{equation}
\underbrace{(\sbx;\sbx) \otimes\dots \otimes (\sbx;\sbx)}_{Q}
=(\, \underbrace{\vphantom{()}\Qsbx}_{Q}\,
;\,\underbrace{\vphantom{()}
\Qsbx}_{Q}\, )\oplus\cdots\,,
\label{brule}
\end{equation}
where the dots represents the direct sum of other irreducible representations.
In particular, the representation $\Qsbx$ being again a short representation under 
$\fpsu(2|2)\ltimes {\mathbb{R}}^{3}$ satisfies the shortening condition in \cite{Beisert3} 
and carries protected central charges.
Furthermore, by considering the multi-magnon boundstates in the
$SU(2)$ spin chain, we can conclude that the most general $Q$-magnon boundstate should be contained in the first term of the decomposition
(\ref{brule}), as such term contains 
a state of highest weight $Q$.  
It should be a straightforward 
but tedious excercise to decompose $(\Qsbx ;\Qsbx)$ into the
irreducible representations of $SU(2)^{4}$, and rewrite the
various terms in the decomposition in terms of the $\N=4$ SYM fields
as we did for the case of $Q=2$. It would also be interesting to
identify these different species of boundstates from 
the poles in their associated scattering matrices \cite{Beisert:2005tm,Beisert:2005fw}.
Even though the classification here does not completely rule out the possibility 
of having boundstates in other irreducible representation at larger $Q$, the states
in $(\Qsbx\,;\Qsbx)$ should be regarded as the minimal set of boundstates in the asymptotic spectrum.

Here we would like to discuss the number of the possible polarisations
for a $Q$-magnon boundstate. 
In decomposing the 
irreducible representations of $SU(2|2)$ into those of the
$SU(2)\times SU(2)$ subgroup, the valid Young tableau
involved should only contain single rows to comply with the usual
rules. As the result the decomposition for irreducible representation of our 
interests terminates after three terms: 
\begin{equation}
\underbrace{\vphantom{()}\Qsbx}_{Q}=(\underbrace{\vphantom{()}\Qbx}_{Q},{\bf
  1})+(\underbrace{\vphantom{()}\Qbx}_{Q-1},\bx)+
(\underbrace{\vphantom{()}\Qbx}
_{Q-2},{\bf1})\,.
\label{Qsbx}
\end{equation}
Simple counting shows that there are $4Q$ states in this
decomposition, and for 
$(\Qsbx\,;\Qsbx)$ which contains all possible polarisations for $Q$-magnon boundstates, there are $(4Q)^{2}=16Q^{2}$ states. This is the
degeneracies for a given boundstate charge $Q$ and it is drastically
different from the number of possible out-going states for $Q$-magnon
scatterings, which goes exponentially with $Q$.
This concludes our discussion on the representation of the 
magnon boundstates. 
\paragraph{}
Having worked out the representation, 
it is rather straightforward to
obtain an exact dispersion relation for the general $Q$-magnon bound
states by 
extending the arguments in \cite{Beisert:2005tm}.
The idea is that, as we discussed earlier, the energy $E$ of the
magnon boundstate should again be the physical central charge $\cC$
carried by the associated irreducible representation $(\Qsbx\,;\Qsbx)$
under the extended 
residual symmetry algebra. 
Recall that this central charge (along with the two extra ones) is
shared between the two $\fsu(2|2)$s in the extended algebra, in
addition, the magnon boundstate transforms under identical short irreducible
representation with respect to each $\fsu(2|2)$. We conclude that it
is sufficient to consider the action of only single $\fsu(2|2)$ (with two extra central charges)
on the boundstate, and treat the components transforming under the
other $\fsu(2|2)$ as the spectators, 
just like the infinite number of background $Z$ fields. 
Moreover, as $\fC$ should commute with other group generators which
relate all $16Q^{2}$ different polarisations for magnon boundstate of
charge $Q$, 
the dispersion relation deduce here would be identical for all of them.
    
Moving onto a $Q$-magnon boundstate which transforms as $\Qsbx$ under $\fpsu(2|2)\ltimes {\mathbb{R}}^{3}$, 
so we have  
\begin{equation}
\underbrace{\vphantom{()}\Qsbx}_{Q}\, ~:\quad 
\kket{\Xi_{Q}}\eq\kket{\xi^{(A_{1}}\xi^{A_{2}}\dots\xi^{A_{Q-1}}\xi^{A_{Q})}}\,,\label{Qbound}
\end{equation}
where we have omitted the infinite number of background $Z$ fields. Here we have also introduced $\xi^{A_{i}}=\{\phi^{a_{i}};\psi^{\al_{i}}\}$ a generalized vector and $A_{i}=\{a_{i},\alpha_{i}\}$ a generalized index for notational conveniences. 
We are interested in the central charge $\bcC_{Q}$ carried by such state, which would in turn give us the required dispersion relation. 
This can be obtained by considering the actions from both sides of the commutator (\ref{comm3}) on the higher tensor representations, and combining with the algebraic relations (\ref{Q-phi})-(\ref{L-phi}).

We shall give our calculational details in the generalized indices $A_{i}$ and only focus on the algebraic structures, the explicit conversion into bosonic and fermionic indices, $a_{i}$ and $\alpha_{i}$ respectively, should be obvious. 
First let us act the LHS of (\ref{comm3}) on $\kket{\Xi_{Q}}$ using (\ref{Q-phi})-(\ref{S-psi}) to obtain
\begin{align}
\sum_{i=1}^{Q}\{ \fQ, \fS \}_{A_{i}}^{B_{i}}\kket{\Xi_{Q}^{C_{i}}}
&=\sum_{i=1}^{Q}\ko{\a_{i}\d_{i}-\bb_{i}\c_{i}}\delta^{C_{i}}_{A_{i}}\kket{\Xi_{Q}^{B_{i}}}
+\sum_{i=1}^{Q}\bb_{i}\c_{i}\delta^{B_{i}}_{A_{i}}\kket{\Xi_{Q}^{C_{i}}}\,.
\label{LHS}
\end{align}
The notation here means that $\{\fQ,\fS\}^{B_{i}}_{A_{i}}$ only acts on the $i$-th fundamental representation in the tensor representation and the superscript $C_{i}$ in $\kket{\Xi_{Q}^{C_{i}}}$ is also for highlighting such fact.

On the other hand, the action of the RHS of (\ref{comm3}) on $\kket{\Xi_{Q}}$ gives.
\begin{equation}
\sum_{i=1}^{Q}\left(\fL+\fR+\fC\right)^{B_{i}}_{A_{i}}\kket{\Xi_{Q}^{C_{i}}}
=\sum_{i=1}^{Q}\left\{\delta^{C_{i}}_{A_{i}}\kket{\Xi_{Q}^{B_{i}}}
+\left(\cC_{i}-\frac{1}{2}\right)\delta^{B_{i}}_{A_{i}}\kket{\Xi_{Q}^{C_{i}}}\right\}\,,
\label{RHS}
\end{equation}
where we have used $\cC_{i}$ to denote the central charge carried by the $\xi^{A_{i}}$, that is $\fC\kket{\xi^{A_{i}}}=\cC_{i}\kket{\xi^{A_{i}}}$.
From (\ref{LHS}) and (\ref{RHS}), we can deduce the closure of the symmetry algebra requires 
\begin{equation}
\a_{i}\d_{i}-\bb_{i}\c_{i}=1\quad \mbox{and}\quad {\cC}_{i}=\ko{\half+\bb_{i}\c_{i}}\,,~~~i=1\,,\dots\,,Q\,,
\end{equation}
which then implies that $\cC_{i}=\frac{1}{2}(\a_{i}\d_{i}+\bb_{i}\c_{i})$. The central charge $\bcC_{Q}$ of $\kket{\Xi_{Q}}$ is given by sum of the individual central charges, so we have
\begin{equation}
\bcC_{Q}=\sum_{i=1}^{Q}\cC_{i}=\f{1}{2}\sum_{i=1}^{Q}\ko{\a_{i}\d_{i}+\bb_{i}\c_{i}}=\f{1}{2}\ko{\A\D+\B\C}\,.
\label{cC}
\end{equation}
This is the central charge of the $Q$-magnon boundstate in terms of $\a_{i}$, $\bb_{i}$, $\c_{i}$ and $\d_{i}$, 
and here we have also introduced $\A,\B,\C$ and $\D$ which should be functions of the spectral parameters for the boundtstates $X^{\pm}$.
To proceed obtaining the explicit expression for $\bcC_{Q}$, we need to work out $\A,\B,\C$ and $\D$ or at least some combinations of them in terms of the magnon boundstate spectral parameters, this is where the two extra central charges $\fP$ and $\fR$ in (\ref{comm4}) and (\ref{comm5}) come in. 
First consider the actions of (\ref{comm4}) and (\ref{comm5}) on $\kket{\Xi_{Q}}$, one can deduce that 
\begin{align}
\fP\kket{\Xi_{Q}}
&=\sum_{i=1}^{Q}\a_{i}\bb_{i}\prod^{Q}_{j=i+1}\exp(-ip_{j})\kket{\Xi_{Q}^{C_{i}}Z^{+}}\,,\label{PQbound}\\
\fK\kket{\Xi_{Q}}
&=\sum_{i=1}^{Q}\c_{i}\d_{i}\prod^{Q}_{j=i+1}\exp(ip_{j})\kket{\Xi_{Q}^{C_{i}}Z^{-}}\,.\label{KQbound}
\end{align}
In deducing (\ref{PQbound}) and (\ref{KQbound}), we have also used the consistency relation 
$\kket{Z^{\pm}\xi^{A_{i}}}=\exp(\mp i p_{i})\kket{\xi^{A_{i}}Z^{\pm}}$ to shift the insertion/removal of $Z$ field to the far right.

Using the expressions for $\a_{i},\bb_{i},\c_{i}$ and $\d_{i}$ in \cite{Beisert:2005tm}, we have $\a_{i}\bb_{i}=\alpha(\exp(-ip_{i})-1)$ and $\c_{i}\d_{i}=\beta(\exp(ip_{i})-1)$ with $\al$ and $\be$ some constants for the time being. 
The two additional central charges carried by the magnon boundstate are given by
\begin{equation}
{\bcP}_{Q}=\alpha\left(\prod_{i=1}^{Q}\exp(-ip_{i})-1\right)=\A\B~~~{\rm{and}}
~~~{\bcK}_{Q}=\beta\left(\prod^{Q}_{i=1}\exp(ip_{i})-1\right)=\C\D\,.\label{2ccharges}
\end{equation} 
The momentum carried by the magnon boundstates should be the sum of constituent momenta, this allows us to write down
\begin{equation}
{\bcP}_{Q}=\A\B=\alpha(e^{-iP}-1)=\alpha\left(\frac{X^{-}}{X^{+}}-1\right)\,,~~~
{\bcK}_{Q}=\C\D=\beta(e^{iP}-1)=\beta\left(\frac{X^{+}}{X^{-}}-1\right)\,,\label{ABCD}
\end{equation}
where $P=\sum_{i=1}^{Q}p_{i}$ is the momentum carried by the $Q$-magnon boundstate.
When we restrict to the physical states which living in $\fpsu(2|2)\ltimes {\mathbb{R}}^{3}$, both of extra central charges should vanish.

Moreover, as the $Q$-magnon boundstates transform in the short representation $\Qsbx$ of $\fpsu(2|2)\ltimes{\mathbb{R}}^{3}$, in terms of its central charges $\bcC_{Q},\bcP_{Q}$ and $\bcK_{Q}$, the shortening condition it obeys is
\begin{equation}
\bcC_{Q}^{2}-\bcP_{Q}\bcK_{Q}=\frac{Q^{2}}{4}\,.\label{shortening2}
\end{equation}
In the light of (\ref{shortening1}) and (\ref{constraint1}),
this should in turn provide a constraint on the boundstate spectral parameters $X^{\pm}$ as
\begin{equation}
X^{+}+\frac{g^{2}}{2X^{+}}-X^{-}-\frac{g^{2}}{2X^{-}}=iQ\,.\label{constraint2}
\end{equation}
This can be guaranteed and reduced correctly to trivial $Q=1$ case if we set 
\begin{equation}
\A\D=\sum^{Q}_{i=1}\a_{i}\d_{i}=-i(X^{+}-X^{-})\,,
\qquad 
\B\C=\sum^{Q}_{i=1}\bb_{i}\c_{i}=i\frac{g^{2}}{2}\left(\frac{1}{X^{+}}-\frac{1}{X^{-}}\right)\,.\label{XpXm}
\end{equation}
Using the explicit expressions for $\a_{i},\bb_{i},\c_{i}$ and $\d_{i}$ in terms of the magnon spectral parameters given in \cite{Beisert:2005tm}, we deduce that
\begin{equation}
X^{+}-X^{-}=\sum^{Q}_{i=1}(x_{i}^{+}-x_{i}^{-})\,,
\qquad 
\frac{1}{X^{+}}-\frac{1}{X^{-}}=\sum^{Q}_{i=1}\left(\frac{1}{x_{i}^{+}}-\frac{1}{x_{i}^{-}}\right)\,.\label{XpXm2}
\end{equation}
Combining (\ref{XpXm2}) with (\ref{ABCD}), they give three constraints on $\{x^{\pm}_{1}\,,\dots\,,x^{\pm}_{Q}\}$ in terms of $X^{\pm}$ which can be satisfied by the combination
\begin{eqnarray}
&&X^{+}=x^{+}_{1}\,,~~~X^{-}=x^{-}_{Q}\,,\label{constraint3}\\
&&x_{i}^{-}=x_{i+1}^{+}\,,~~~i=1\,,\dots\,, Q-1\label{constraint4}\,.
\end{eqnarray}   
The equation (\ref{constraint4}) is identical to the multiplet splitting condition given earlier (\ref{splitting2}), as the ``super-symmetric'' representation $\Qsbx$ can only arise from the decomposition of general $Q$-magnon tensor product after (\ref{splitting2}) is imposed.

From (\ref{ABCD}) and (\ref{shortening2}) or (\ref{constraint2}), we can also deduce $\bcC_{Q}$ for the magnon boundstate
\begin{equation}
\bcC_{Q}=\f{1}{2}\sqrt{\ko{\A\D-\B\C}^{2}+4\A\B\C\D}
=\f{1}{2}\sqrt{Q^{2}+16\al \be \sin^{2}\ko{\f{P}{2}}}\,.
\end{equation}
The product $\al\be$ is in general a function of the 't Hooft coupling $\lambda$. For the case of single magnon, it has been set to $\al\be=\lambda/16\pi^{2}$ by considering the BMN limit \cite{Berenstein:2002jq}, this dependence should interpolate to case of $Q> 1$, and indeed one can confirm that for example by considering the Frolov-Tseytlin limit \cite{Frolov:2003xy} as in \cite{Dorey:2006dq}. In any case, we deduce that the dispersion relation for the magnon boundstate from the group theoretical means is 
\begin{equation}
E=\Delta-J_{1}\eq 2\bcC_{Q}=\sqrt{Q^{2}+\f{\lambda}{\pi^{2}}\sin^{2}\ko{\f{P}{2}}}\,.\label{kdisprel}
\end{equation}
This formula reduces the one proposed in \cite{Dorey:2006dq} for single magnon boundstate of charge $Q=1$, with $\Delta-J_{1}$ coincides with (\ref{Magnondispersion}). It is also important to note that, as discussed earlier, there will be $16Q^{2}$-fold degeneracies which correspond to the all possible polarisations of a $Q$-magnon boundstate, all share the same dispersion relation (\ref{Magnondispersion}).

We would also like to make a comment on the case when there are more than one boundstate in the asymptotic spin chain, namely a state of the form $\kket{\Xi_{Q_{1}}\dots\Xi_{Q_{M}}}$.
Here $M$ is the number of the boundstates each of which are well-separated, and $Q_{k}$ is the number of constituent magnons in the $k$-th boundstate.
In this case the dispersion relation (\ref{kdisprel}) is simply generalised to give
\begin{equation}
E=\Delta-J_{1}\eq \sum_{k=1}^{M}2\bcC_{Q_{k}}\quad \mbox{with}\quad  
\bcC_{Q_{k}}\eq \f{1}{2} \sqrt{Q_{k}^{2}+\f{\lambda}{\pi^{2}}\sin^{2}\ko{\f{P_{k}}{2}}}\,,
\end{equation}
where $P_{k}$ is the total momentum of the $k$-th boundstate in the
asymptotic spin chain.
\paragraph{}
In this paper we have described the infinite tower of BPS boundstates 
appearing in the asymptotic spectrum of the ${\cal N}=4$ spin chain and 
identified the corresponding representation of supersymmetry in which they 
transform. 
As these are short representations we expect that these states are present 
for all values of the 't Hooft coupling, $\lambda$. Indeed, as discussed in 
\cite{Dorey:2006dq,Chen:2006ge}, the representatives of the 
boundstate multiplets lying in a given $SU(2)$ sector are directly visible 
both in one-loop gauge theory and in semiclassical string theory which 
correspond to small and large $\lambda$ respectively. An obvious question is 
whether additional asymptotic states are also present. At this point we cannot 
rule out the possibility that some of the additional short representations, 
which appear in the tensor product of 
bifundamentals when the shortening condition is obeyed, 
also correspond to BPS boundstates in the spectrum. 
However, the representations 
which can occur are certainly constrained by the known absence of boundstates 
in the remaining rank one sectors. In particular this rules out additional 
boundstates with $Q=2$.  

In closing, we should note that there are two classes of states which we have 
not included in our discussion. First, the semiclassical string theory 
analysis of \cite{Hofman:2006xt} suggests the presence of an infinite tower 
of neutral boundstates appearing as poles in the two-magnon S-matrix. 
These poles should appear at values of the kinematic variables which do not 
satisfy the shortenting condition. In fact, for such 
generic values of the momenta the tensor product of two bifundamentals 
actually consists of a single irreducible long multiplet \cite{Beisert3}. 
Each of the neutral boundstates of \cite{Hofman:2006xt} must therefore fill 
out such a multiplet. As the energies of these states are not protected, their 
behaviour away from the region of large 't Hooft coupling is still unclear. 
Finally we recall that the ${\cal N}=4$ spin chain also contains a 
singlet state of zero energy \cite{Beisert:2005tm}. However, in a crossing 
invariant theory, this state is indistinguishable from the vacuum.           
\paragraph{}
The authors are grateful to Niklas Beisert and Juan Maldacena 
for useful discussions. 
HYC is supported by a Benefactors' scholarship from St.\,John's
College, Cambridge. ND is supported by a PPARC Senior Fellowship. 
HYC thanks University of Michigan, Ann Arbor,
University of Wisconsin, Madison and Brown University, for their hospitalities during 
the final stages of the preparation.
KO is very grateful to University of Cambridge, 
Centre for Mathematical Sciences, where part of the work was done, 
for its warm hospitality.

\subsubsection*{Appendix: Decomposition of \bmt{\N=4} Fields into \bmt{SU(2)^{4}} Representations\label{app:rep}}

Here we list all possible tensor decompositions between two $\N=4$ SYM excitations for readers' convenience.
\begin{alignat}{3}
\Phi_{i}&\otimes \Phi_{j}\quad &:\quad &(\bx,{\bf 1};\bx,{\bf 1})\otimes (\bx,{\bf 1};\bx,{\bf 1})\cr
&&&\quad \quad =(\twobx,{\bf 1};\twobx,{\bf 1})
\oplus(\twobx,{\bf 1};{\bf 1},{\bf 1})
\oplus({\bf 1},{\bf 1};\twobx,{\bf 1})
\oplus({\bf 1},{\bf 1};{\bf 1},{\bf 1})\,,\\
D_{a}&\otimes D_{b}\quad &:\quad &({\bf 1},\bx;{\bf 1},\bx)\otimes ({\bf 1},\bx;{\bf 1},\bx)\cr
&&&\quad \quad =({\bf 1},\twobx;{\bf 1},\twobx)
\oplus({\bf 1},\twobx;{\bf 1},{\bf 1})
\oplus({\bf 1},{\bf 1};{\bf 1},\twobx)
\oplus({\bf 1},{\bf 1};{\bf 1},{\bf 1})\,,\\
D_{a}&\otimes \Phi_{i}\quad &:\quad &({\bf 1},\bx;{\bf 1},\bx)\otimes (\bx,{\bf 1};\bx,{\bf 1})
=(\bx,\bx;\bx,\bx)\,,\\
\Psi_{{\dot\al}\dot\be}&\otimes \Psi_{\ga{\delta}}\quad &:\quad &({\bf 1},\bx;\bx,{\bf 1})\otimes (\bx,{\bf 1};{\bf 1},\bx)=(\bx,\bx;\bx,\bx)\,,\\
\Psi_{\al{\be}}&\otimes \Psi_{\ga{\delta}}\quad &:\quad &(\bx,{\bf 1};{\bf 1},\bx)\otimes (\bx,{\bf 1};{\bf 1},\bx)\cr
&&&\quad \quad =(\twobx,{\bf 1};{\bf 1},\twobx)
\oplus(\twobx,{\bf 1};{\bf 1},{\bf 1})
\oplus({\bf 1},{\bf 1};{\bf 1},\twobx)
\oplus({\bf 1},{\bf 1};{\bf 1},{\bf 1})\,,\\
\Psi_{{\dot\al}\dot\be}&\otimes \Psi_{{\dot\ga}\dot\delta}\quad &:\quad &({\bf 1},\bx;\bx,{\bf 1})\otimes ({\bf 1},\bx;\bx,{\bf 1})\cr
&&&\quad \quad =({\bf 1},\twobx;\twobx,{\bf 1})
\oplus({\bf 1},\twobx;{\bf 1},{\bf 1})
\oplus({\bf 1},{\bf 1};\twobx,{\bf 1})
\oplus({\bf 1},{\bf 1};{\bf 1},{\bf 1})\,,\\
\Phi_{i}&\otimes \Psi_{\al{\be}}\quad &:\quad &(\bx,{\bf 1};\bx,{\bf 1})\otimes (\bx,{\bf 1};{\bf 1},\bx)
=({\bf 1},{\bf 1};\bx,\bx)
\oplus(\twobx,{\bf 1};\bx,\bx)\,,\\
\Phi_{i}&\otimes \Psi_{{\dot\al}\dot\be}\quad &:\quad &(\bx,{\bf 1};\bx,{\bf 1})\otimes ({\bf 1},\bx;\bx,{\bf 1})
=(\bx,\bx;{\bf 1},{\bf 1})
\oplus(\bx,\bx;\twobx,{\bf 1})\,,\\
D_{a}&\otimes \Psi_{\al{\be}}\quad &:\quad &({\bf 1},\bx;{\bf 1},\bx)\otimes (\bx,{\bf 1};{\bf 1},\bx)
=(\bx,\bx;{\bf 1},{\bf 1})
\oplus(\bx,\bx;{\bf 1},\twobx)\,,\\
D_{a}&\otimes \Psi_{{\dot\al}\dot\be}\quad &:\quad &({\bf 1},\bx;{\bf 1},\bx)\otimes ({\bf 1},\bx;\bx,{\bf 1})
=({\bf 1},{\bf 1};\bx,\bx)
\oplus({\bf 1},\twobx;\bx,\bx)\,.
\end{alignat}

%%%%%%%%%%%%%%%%%%%%%%%%%
%%%%%%%%%%%%%%%%%%%%%%%%%

\end{document}